# Evolutionary Optimisation Methods for Template Based Image Registration

*Lukasz A Machowski, Tshilidzi Marwala*

School of Electrical and Information Engineering
University of Witwatersrand, Johannesburg, South Africa.

l.machowski@ee.wits.ac.za  t.marwala@ee.wits.ac.za

## Abstract

This paper investigates the use of evolutionary optimisation techniques to register a template with a scene image. An error function is created to measure the correspondence of the template to the image. The problem presented here is to optimise the horizontal, vertical and scaling parameters that register the template with the scene. The Genetic Algorithm, Simulated Annealing and Particle Swarm Optimisations are compared to a Nelder-Mead Simplex optimisation with starting points chosen in a pre-processing stage. The paper investigates the precision and accuracy of each method and shows that all four methods perform favourably for image registration. SA is the most precise, GA is the most accurate. PSO is a good mix of both and the Simplex method returns local minima the most. A pre-processing stage should be investigated for the evolutionary methods in order to improve performance. Discrete versions of the optimisation methods should be investigated to further improve computational performance.

## 1. Introduction

Image registration has great practical application in the field of computer vision, medicine, remote sensing and image watermarking [1][2]. Being able to determine how best the template image fits into the scene poses several problems that have to be overcome. The registration process may involve shifting, scaling, rotation, perspective projection or other non-linear transformations. The shear number of possible transformations makes it difficult to automate the process and usually requires a person to verify the results manually. This paper presents findings on the use of evolutionary optimisation methods for automating the template matching of 2-dimensional intensity images.

### 1.1 Image Registration

Image registration is the process by which a template is oriented in such a way as to match an entire, or a portion of, a given scene [1][2]. The template is transformed in such a way as to match the scene as closely as possible.

There are four main steps required for registration of an image. These are feature detection, feature matching, transform model estimation and image transformation [1]. Feature based detection makes it easier to determine the orientation of the template with respect to the scene. Area based detection methods are much more computationally expensive due to the amount of data that needs to be processed. Since the area based detection methods depend on the appearance of the images, they are intolerant of changes in illumination and ambient conditions [1][2]. The feature based detection methods do not suffer from this but it is more difficult to automatically extract the features for any general image. It is common to combine the advantages from both methods to form a hybrid approach to the registration process [2].

Correlation-like methods are typically used for area-based detection methods where a correlation surface is calculated for the template and the maximum point is found and interpreted as the best fit for the template [1]. This method is adversely affected by self similarity in the image and it is characterised by high computational complexity. It also does not allow much variance in template rotation or other more complex transformations. This approach, is still however attractive for real-time object tracking [1][3].

An alternative to cross correlation is to use optimisation to find the best fit for the template in the scene [4]. The advantage of this approach is that one can apply more complex transformations to the templates, and thus make the method robust when compared to cross correlation. This method also requires less computation because the entire correlation surface does not have to be determined.

In this paper, we investigate the use of a Genetic Algorithm (GA), Simulated Annealing (SA) and Particle Swarm Optimisation (PSO) to register a template with a given scene. These methods are also compared to the Nelder-Mead Simplex method. For simplicity, only three transformation parameters are defined. These are horizontal translation, vertical translation and uniform scaling.

## 1.2 Evolutionary Optimisation

The term "evolutionary" refers to the fact that the optimum solution gradually evolves from a population of individuals that share information and have group dynamics [5]. This is in contrast to the non-evolutionary or classical optimisation methods which always try to travel in the best direction. Typically, the evolutionary concept is linked with GA alone but in this paper, we group GA, SA and PSO into the subset. All evolutionary optimisation methods have the following operations [5]:

- Evaluation
- Selection
- Alteration

An initial population of individuals is initialised, covering the parameter space and the objective function is evaluated for each individual. From this data, a subset of individuals is selected and altered to form new individuals. The degree to which each of these operations is performed in GA, SA and PSO varies from algorithm to algorithm.

To find the optimal registration parameters for template matching, it is important to construct a multivariate cost function that represents how well the template matches the scene [6]. The traditional techniques for optimisation make use of the objective function value, first derivative or its second derivative [6][7]. The general approach for all non-evolutionary optimisation methods is to select an initial guess for the registration parameters and travel in a direction as to improve the objective function. Once a suitable direction is found, it is possible to make either fixed or varying successive steps towards the local optimum.

Evolutionary Optimisation methods, however, do not make use of any other information but the objective function values themselves. This eliminates the evaluation of gradients which may be expensive and misleading for image registration. Evolutionary methods typically sample the search space significantly more than the non-evolutionary techniques but this improves the probability of the algorithm finding the global optimum point. The evolutionary algorithms are typically based on the processes which occur in the natural world, such as genetics, the swarming behaviour of bees and the annealing of metals. Various data structures are used to simulate these [5].

A brief review of the methods suitable for image registration is given below:

### 1.2.1 Simplex Method

This is a non-evolutionary (classical) method. A simplex is a geometric figure that has one more vertex than the number of dimensions in the parameter space (a triangle in two dimensions, as shown in Figure 1). The objective function is sampled at each vertex and the one that has the worst value gets removed from the simplex. A new vertex is then created by reflecting the simplex about the remaining points. Depending on whether the fitness of the new point improves or not, the simplex is expanded or contracted to look for a more precise solution. In this manner, the algorithm steps its way towards the local optimum point. This method is relatively robust when used for discontinuous objective functions [8].

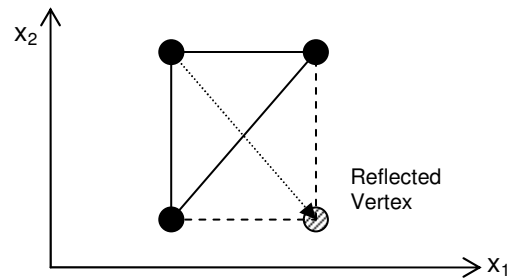

**Figure 1 A simplex in two dimensions, showing a reflection.**

### 1.2.2 Simulated Annealing

Simulated Annealing is a Monte Carlo Technique [5] and is based on the analogy of metals cooling slowly to form a crystalline structure with low energy.

The SA introduces a probability of acceptance of a new sample point for a hill-climbing process. The probability of accepting the new point is based on a control variable called the temperature. The higher the temperature, the more likely it is for a worse point to be accepted. This allows the algorithm to escape from local optima [5][9]. This algorithm is commonly referred to as Metropolis after its founder.

The SA technique used in this paper modifies the simplex method described above to allow the simplex to accept a worse vertex with a probability distribution that is based on the temperature. If the change in energy is negative (we have a better point) then the new point will always be chosen. If the change in energy is positive (the point is worse) then the probability of accepting it is given by:

$$p = e^{\frac{-\Delta E}{k_B \cdot T}} \quad (1)$$

where $\Delta E$ is the change in energy, $k_B$ is Boltzmann's constant and $T$ is the current temperature. The cooling schedule (how many iterations to spend at each temperature) is an important factor in the success of the algorithm.

### 1.2.3 Genetic Algorithm

Based on the theory of genetics, the GA encodes each individual in the population with a chromosome [5][10]. This encoding represents the parameters for the objective function being optimised. There are several different techniques for encoding parameters, performing the selection, and the alteration stages of the algorithm. The alteration stage is separated into Crossover and Mutation. The method used in this paper selects a random sample of parents from the population with a specified probability. An arithmetic crossover is then performed on these individuals which creates children based on a linear interpolation of the two parents. This is shown in Figure 2.

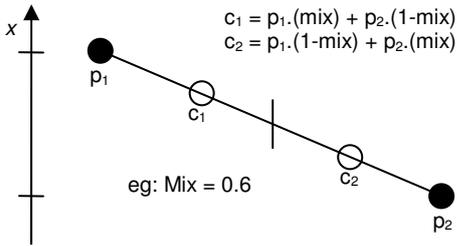

**Figure 2 One Dimensional Arithmetic crossover operator.**

A multi-non-uniform mutation is performed which modifies the parent parameters with a binomial distribution which narrows as the number of generations gets larger. More details of the individual techniques are given in [5].

### 1.2.4 Particle Swarm Optimisation

PSO is based on the swarming behaviour of bees, flocking of birds, schooling of fish and social relations of humans [11]. A population of particles is randomly initialised within the parameter space and each one is given an initial velocity. At each iteration of the algorithm, the particle position is updated and a new velocity is calculated, taking the best position for the particle and the group into account. There are different ways of grouping the particles together. The method used in this paper creates a social grouping where each particle has *n* logical neighbours (referenced by adjacent index numbers). This means that particles can be neighbours even though they are not close to each other spatially. This method also tends to produce better global exploration by the particles since there are many more attractors. The velocity of each particle is calculated as:

$$v_d(t+1) = \begin{pmatrix} \alpha.v(t) \\ + \beta_i.(p_{i,d} - x_d(t)) \\ + \beta_g.(p_{g,d} - x_d(t)) \end{pmatrix} \quad (2)$$

where $v_d(t+1)$ is the next velocity for particle d, α is an acceleration constant, $\beta_i$ is an attraction constant for the individual best position, $\beta_g$ is an attraction constant for the group best position, $p_{i,d}$ is the best individual position for particle d, $p_{i,g}$ is the best group position for particle d's neighbourhood and $x_d(t)$ is the current position for particle d. The position of each particle is then calculated for each iteration as:

$$x_d(t+1) = x_d(t) + v_d(t+1) \quad (3)$$

The various components that make up the velocity for each particle are shown in Figure 3.

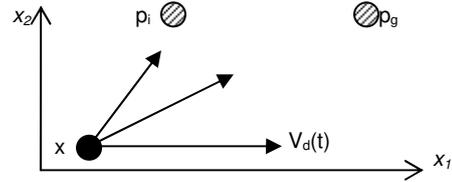

**Figure 3 Various components making up the new velocity.**

## 2. Method

### 2.1 Objective Function

In order to perform optimisation, it is necessary to define an objective function that captures the essence of the problem at hand. In image registration, one wants to maximise correspondence between the scene image and the template posed at its current position. The correspondence can be measured as the sum-squared difference between the intensities of overlapping pixels. This can be expressed as an error function where a value of zero represents a perfect match. The parameters to be optimised in this problem are horizontal translation (x), vertical translation (y) and uniform scaling (s). The objective function used when there are overlapping pixels between the template and the scene image is given by:

$$error = \sum (A-T)^2 / numel(A) \quad (3)$$

where A is the scene image, T is the template, the $^2$ refers to element-wise squaring and the summation is over each element of the resultant matrix. This error is then normalised with the number of pixels that are overlapping between both images.

It is also important to interpolate sub-pixel values for the optimisation algorithm to be able to function correctly. This allows the traditional algorithms to be run unchanged and also allows it to be compared to other general optimisation algorithms.

It is necessary to penalise the error function when there are template pixels that do not lie within the image. This error component is added to the existing value calculated above. The penalised error then becomes:

$$error_p = error + OutPixels \times c \quad (4)$$

where error is from equation (3), OutPixels is the number of template pixels that do not lie in the image, and c is a penalisation constant which should be large (~1000). This penalisation has the effect of constraining the x and y parameters back into range when the images no longer overlap. It is necessary to hard-limit the scale parameter because the optimisation algorithms might try ridiculously high values which require extremely large amounts of memory. Very seldom does a template match a scene at very high scaling values. Similarly, if the template gets scaled to one pixel in size, then a fit can be found nearly anywhere in the scene.

### 2.2 Test Image

The test image used is the familiar picture of Lena, which has a good mix of various image features and provides several local minima for registration. The template is a cut-out of Lena's face and is shown in Figure 4. The image scene is 256x256 pixels and the template is 170x138 pixels taken from the coordinates (151.5, 151.5) with a scale of 2.0. This means that the global optimum for our objective function is at the coordinates (151.5, 151.5, 0.5) with an error value of 0.

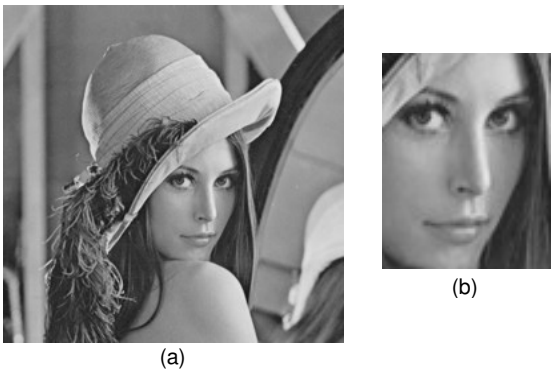

**Figure 4 a) Test Image b) Template for registration.**

### 2.3 Optimisation

This section describes implementation details for each of the optimisation algorithms. All routines are implemented in Matlab R13. The implementation details for each algorithm are given below:

#### 2.3.1 Simplex Method

Matlab's Optimisation Toolbox is used to perform the Nelder-Mead simplex optimisation [8]. The function used is *fminsearch.m*. A simple pre-processing algorithm is used to choose initial starting points.

#### 2.3.2 Simulated Annealing

The SA algorithm used in this paper is based on the code given in [12] and is extended to include restarts. The code is abstracted into a higher level in order to take advantage of Matlab's matrix arithmetic capabilities. The following is a high level description of the algorithm used:

```
x = RandomStartingSimplex();
y = EvaluateSimplex(S);
for each temperature in cooling schedule:
   for number of iterations:
      yFluc = AddFluctuation(y);
      sort(yFluc);
      ReflectSimplex;
      If better than best then
         ExpandSimplex(x)
      Else if worse than 2nd highest:
         ContractSimplex(x);
         If still bad then:
            ContractOtherVerices(x);
         End
      End
      If SimplexIsStuck() then:
         % Restart, keeping best point:
         x = RandomStartingSimplex();
         KeepBestPoint(x);
      End
   End for each iteration
End for each temperature
```

The restart allows the algorithm to oscillate at a local minimum for only a limited number of iterations. After this, it is restarted with the best point as one of the vertices. This behaviour is justified because it is recommended in [12] that the algorithm be re-run in the same fashion once a solution is found. This merely allows the simplex to further explore the parameter space.

#### 2.3.3 Genetic Algorithm

The Genetic Optimization Toolbox (GAOT) [13] is used for the implementation of the GA. This is an extensive toolbox with many functions for the encoding, selection, crossover and mutation operators. The following operators were used for the optimisation:

- *normGeomSelect.m*, with the probability of selecting the best, set to 0.6.
- *arithXover.m*, with 10 crossovers per generation
- *multiNonUnifMutation.m*, with 20 mutations per generation.

The particular selection operator used gives a good mix of exploration and precision. The arithmetic crossover is very useful in this problem since it assists in finding more precise parameters for the objective function. It is important to use the given mutation operator so that a sufficient amount of exploration occurs. The nature of the image registration problem creates an objective function with many local minima so it is important to mutate out of these valleys. Since the GAOT maximises an objective function, we merely multiply our original objective function by a factor of -1 to perform the minimisation.

### 2.3.4 Particle Swarm Optimisation

After evaluating the performance of a free PSO toolbox, and getting poor results, it was decided to write a custom PSO routine based on the method described in the previous section. The high level description of the algorithm is given below:

```
Swarm = CreateRandomSwarm();
While we have more iterations to go:
   EvaluateObjectiveFunction(Swarm);
   If Swarm.BestValue<GlobalBestValue then
      GlobalBestValue = Swarm.BestValue;
      GlobalBestPosition=Swarm.BestPosition;
   End
   UpdateIndividualAndGroupBestValues(Swarm);
   CalculateParticleVelocities(Swarm);
   UpdateParticlePositions(Swarm);
End While loop
Output GlobalBestPosition;
```

The initial swarm is created with random particles between the bounds of the parameter space. A record is kept of each particle's best value that it has sampled. Similarly, a group-best is maintained for a social neighbourhood size of 3 particles to either side of the current particle. Modulo indexing is used. The α parameter (described in the previous section) is set to 0.99 so that the swarm does not become unstable and diverge. Both β parameters are set to 0.01 in order that the particles approach the best locations gradually. This samples the objective function many times along the trajectory of the particle.

## 3. Results

### 3.1 Optimisation

In order to be able to compare the algorithms, the maximum number of function evaluations for each method is set to 1000. The algorithms are stopped as close to this value as possible (since the number of function evaluations may vary from run to run). The amount of time taken by each algorithm is not a good measure of its performance in this case because the amount of processing in the objective function is highly dependent on the parameter values being sampled. All four algorithms described in this paper ultimately approach the global optimum so a suitable measure for their performance is to investigate their precision and accuracy. Each algorithm is run 50 times and an accuracy histogram is calculated for how close and how consistently the algorithm reached the global optimum of (151.5, 151.5, 0.5). The distance is measured geometrically by the following equation:

$$d_i = \sqrt{(x_G - x_i)^2 + (y_G - y_i)^2 + (s_G - s_i)^2} \quad (5)$$

where $d_i$ is the distance for the $i^{th}$ run, $(x_G, y_G, s_G)$ are the global optimum parameters and $(x_i, y_i, s_i)$ are the optimum parameters as calculated in the $i^{th}$ run. The histogram is divided into 10 equal bins so that the last bin has a distance which is 10 units away from the global optimum.

The accuracy histograms for the various methods are given in Figure 5 and the results are analysed below. Higher counts towards bin zero are better.

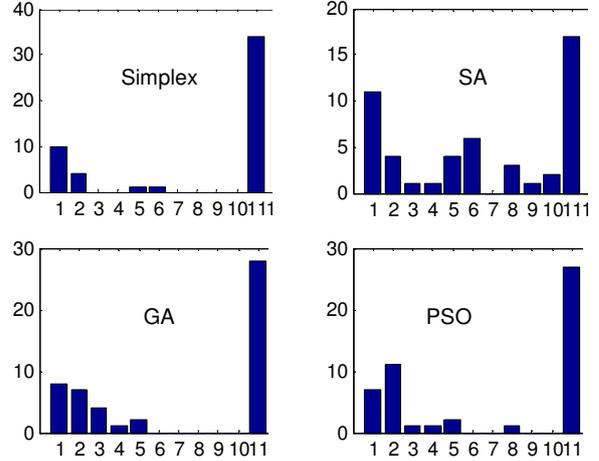

**Figure 5 Accuracy Histograms.**

### 3.1.1 Simplex Method

This method is especially robust for functions that have peculiar gradients or that are discontinuous, as is common in image registration. The method of travel by the simplex acts as a pseudo-gradient that plays a similar role as in the gradient methods. The success of this algorithm is attributed to the pre-processing stage of the algorithm which selects suitable starting points for the Nelder-Mead optimisation. The histogram shows that when the algorithm is near the global optimum, it is reasonably precise. It does however, find local minima quiet often.

### 3.1.2 Simulated Annealing

It is expected that this algorithm performs better than the Simplex Method described above due to the Metropolis method that it employs. It was found that the performance of the algorithm is significantly improved by restarting when the simplex becomes stuck. Without this behaviour, the algorithm tries oscillatory values which waste precious computation time since the samples do not advance the simplex at all. The histogram shows that the algorithm is the least accurate (repeatable) algorithm of the four but is precise (finds values near to the global optimum many times).

### 3.1.3 Genetic Algorithm

The performance of the GA can be attributed to the number of mutations that take place. This is important for image registration because the objective function has many local optima that vary in shallowness. It is more important for the algorithm to explore the parameter space than it is to improve the precision (using the arithmetic crossover). It was also observed that the mutation rate is more important than the initial population size (which randomly explores the parameter space), since sufficient mutations will explore the space more wisely. The histogram shows a clear binomial distribution near the global optimum. This shows that the GA produces reasonably accurate (consistent) results.

### 3.1.4 Particle Swarm Optimisation

The PSO produces acceptable results because of the group dynamics in the system. The social groups promote exploration of the search space while the individual best position lets each particle improve its precision. Good parameter values were found by investigating what effect they have on the swarm behaviour and then tweaking the values to suite the problem domain. The behaviour of the swarm is predictably based on the algorithm parameters so it is relatively easy to infer good parameter values by watching how the particles swarm in the objective function. The histogram shows that the algorithm finds precise results but not always accurately.

### 3.2 Other Sample Data

The algorithms managed to register the templates of distorted and noisy images of Lena's face to the image. The various modifications to the templates that were made include, Gaussian blurring, the addition of Gaussian noise, vortex rotation, smudging and non-uniform stretching. The characteristics and accuracy histograms for each algorithm remain relatively consistent with the results described in the previous section.

## 4. Recommendations

The current objective function makes use of sub-pixel sampling to obtain a continuous parameter space. This is very computationally intensive and unnecessary for certain parameters in image registration such as x and y coordinates. The algorithms presented above should be modified so that the user can specify that certain parameters may vary discretely whilst others continuously. This should reduce the amount of time taken for each function evaluation. Another improvement to the algorithms would be to select regions of interest that are likely to contain the template. This requires looking for higher level features in the image first and then creating bounds for the optimisation algorithms.

## 5. Conclusion

This paper introduced evolutionary optimisation methods and contextualised them in the image registration field. It was found that the algorithms perform well in different aspects of the image registration process. The parameters for each method need to be tuned to suit the given image but the behaviour of the methods presented in this paper are intuitive and insight into how to modify parameters can easily be gained by simulating a few initial runs. These evolutionary optimisation methods were compared to the non-evolutionary Nelder-Mead simplex method with starting points selected by doing higher level pre-processing. SA returns the most precise results, while GA returns the most accurate results. PSO is in between these two and it is followed by the Simplex method. Pre-processing should be investigated for the evolutionary methods.